\documentclass[aps,prl,floatfix,twocolumn]{revtex4}
\usepackage{graphicx}

\begin{document}

\title{A Comparison of the Attempts of Quantum Discord and Quantum Entanglement to Capture Quantum Correlations
\vspace{-2mm}}

\author{Asma Al-Qasimi and Daniel F. V. James}
\affiliation{Department of Physics and Centre for Quantum Information and Quantum Control, University of Toronto, Toronto ON M5S 1A7, Canada.
\vspace{-2mm}}

%\renewcommand{\baselinestretch}{2.0}
%\doublespacing

\begin{abstract} \noindent
Measurements of Quantum Systems disturb their states. To quantify this non-classical characteristic, Zurek and Ollivier \cite{Zurek} introduced the \emph{quantum discord}, a quantum correlation which can be nonzero even when entanglement in the system is zero. Discord has aroused great interest as a resource that is more robust against the effects of decoherence and offers exponential speed up of certain computational algorithms. Here, we study general two-level bipartite systems and give general results on the relationship between discord, entanglement, and linear entropy, and identify the states for which discord takes a maximal value for a given entropy or entanglement, thus placing strong bounds on entanglement-discord and entropy-discord relations. We find out that although discord and entanglement are identical for pure states, they differ when generalized to mixed states as a result of the difference in the method of generalization.
\end{abstract}

\maketitle

Since the emergence of Quantum Mechanics at the beginning of the twentieth century, physicists have been intrigued and puzzled by its interpretation and consequences. Charactersitics that distinguish quantum from classical systems have been investigated extensively. To be able to study quantum correlations in a system, it is important to quantify them. Although this is a challenging task for multi-component quantum systems, progress has been made in the case of two-level bipartite quantum systems. One method suitable for pure states \cite{Wootters2} involves calculating the entropy of the reduced density matrix of the system, also known as the Entanglement of Formation (EoF). To extend this concept to mixed states, Wootters \cite{Wootters} defined it to be equal to the weighted sum of the entanglement of the pure states involved in the decomposition of the mixed state, minimized over all decompositions. The restriction enforced by this minimization places a bound on the entanglement for mixed states; indeed, when a certain level of disorder of the state is attained, it is known that entanglement must disappear \cite{CavesEtAl,Munro}. It is, therefore, not surprising that for some systems as they reach a certain level of mixture, the entanglement is completely lost, a phenomena which in the study of state dynamics is commonly known as Entanglement Sudden Death (ESD) \cite{YuEberly}.

Another approach to capture quantum correlations was taken by Zurek and Ollivier\cite{Zurek}, where they used the fact that measurement of quantum systems, unlike classical systems, disturbs their state. To quantify the correlations based on this idea, one looks at the mutual information function $I(\hat{\rho})$, where $\hat{\rho}$ is the density matrix describing the state of the whole system. Given a system C composed of two subsystems A and B, $I(\hat{\rho})$ tells how much information one can obtain about system A if the state of system B is known (and vice versa). The correlations between the two subsystems can be classical and/or quantum. However, if the correlations are quantum in nature, then calculating $I(\hat{\rho})$ \textit{after} a measurement is performed on one of the subsystems (say B) yields a \textit{different} result to that caluculated \textit{before} the measurement is perfomed. This disagreement is the basis for defining discord; the definition is finalized after optimizing over all possible measurement bases. One of the major reasons in the aroused interest \cite{Caves,White,Datta,Vedral,TQ,XQ,Ferraro,Giorda,Datta2,NMRQ,Vedral2,Datta3,2NQ,SDQ} in this novel correlation is that as was shown \cite{Zurek} even when entanglement is zero in a system, discord can still be finite. This led to the hope that using discord instead of entanglement as a resource, in fields like quantum computation, can lead to more efficient computations. In \cite{Caves}, discord was characterized in the DQC1 (Deterministic quantum computation with one bit) model \cite{Knill}, calling for an experimental verifications of its powers, which was demonstrated in \cite{White}.

The mutual information function is given by:

\begin{equation}
I(\hat{\rho})=S(\hat{\rho}^{A})+S(\hat{\rho}^{B})-S(\hat{\rho}),
\label{Irho}
\end{equation}

\noindent where $\hat{\rho}^{i}$ is the reduced-density matrix of subsystem i and $S\left(\hat{\rho}\right)=-\rm{Tr}\left\{\hat{\rho} \ \rm{log}_{2}\hat{\rho}\right\}$ \cite{vonN}. Then, defining all set of projectors on B by $\{\hat{B}_{k}\}$, the measurement-induced mutual information function for each of these sets takes the following form:

\begin{equation}
I\left(\hat{\rho}\left|\right.\left\{B_{k}\right\}\right)=S\left(\hat{\rho}^{A}\right)-\sum_{k}p_{k}S\left(\hat{\rho}_{k}\right),
\label{Irhok}
\end{equation}

\noindent where $\hat{\rho}_{k}$ is the density matrix of the system after $B_{k}$ is applied on B, $k\in\left\{1,2\right\}$ and $p_{k}=\rm{Tr}\{\left(\hat{I}\otimes \hat{B}_{k}\right)\hat{\rho}\left(\hat{I}\otimes \hat{B}_{k}\right)\}$. To obtain the final form for the measurement-induced density matrix, Eq.(\ref{Irhok}) is maximized over all possible $\{\hat{B}_{k}\}$ to obtain the following expression for discord:

\begin{equation}
Q\left(\hat{\rho}\right)= I(\hat{\rho})-\displaystyle{\mathop{\mbox{max}}_{{{\left\{B_{k}\right\}}}}} \left\{S\left(\hat{\rho}^{A}\right)-\sum_{k}p_{k}S\left(\hat{\rho}_{k}\right)\right\}.
\label{Crhok}
\end{equation}

\noindent To compare discord and entanglement, we compute the EoF and discord for pure states in two-level bipartite systems $\left|\psi\right\rangle=a\left|00\right\rangle+b\left|01\right\rangle+c\left|10\right\rangle+d\left|11\right\rangle$, where $\left|a\right|^{2}+\left|b\right|^{2}+\left|c\right|^{2}+\left|d\right|^{2}=1$. It is convenient to use the Schmidt decomposition \cite{Schmidt} to write the state as $\left|\psi\right\rangle=\lambda\left|1_{A}\right\rangle\left|1_{B}\right\rangle+\left(1-\lambda\right)\left|2_{A}\right\rangle\left|2_{B}\right\rangle$, where $\lambda$ and $\left(1-\lambda\right)$ are the eigenvalues of the reduced density matrices, and $\left|1_{i}\right\rangle$ and $\left|2_{i}\right\rangle$ are the corresponding eigenvectors of the reduced density matrix of subsytem i. Using local unitary operations, which do not affect the quantum correlations present in the system, one can show that this state is equivalent to $\left|\psi\right\rangle=\lambda\left|00\right\rangle+\left(1-\lambda\right)\left|11\right\rangle$. In this case, the EoF as well as discord, which can be computed analytically, are found to be identical and are given by: $E\left(\hat{\rho}\right)=Q\left(\hat{\rho}\right)=h(\lambda)$, where $h(x)=-x\ \rm{log}_{2}\ x-\left(1-x\right)\rm{log}_{2}\left(1-x\right)$. Therefore, discord and entanglement of formation amount to the same set of correlations in the case of pure states \cite{Datta2}. In the mixed state case, there is no explicit analytic expression for discord. The most general analytic expression so far was presented in a very interesting paper by Luo \cite{Luo} for the mixed states with maximally mixed marginals (MMMS) (i.e, the reduced density matrices $\hat{\rho}^{A}$ and $\hat{\rho}^{B}$ are both maximally mixed).

In this Letter, we look at the relationship between discord, entanglement and linear entropy to investigate the connection between these quantities. We perform the study on the most general density matrices. Since we lack an analytic expression for discord for general two-qubit states, the heart of the work is numeric in nature, involving optimization over all possible measurements that can be performed on one of the subsystems under study. 

Going back to system C described above, with A and B each being two-level quantum systems, we parametrize all the possible measurements that can be performed on B by two variables: $\theta$ and $\phi$. Each complete set of possible measurements, which is composed of two elements, is defined as follows:

\begin{eqnarray}
\hat{B}_{1}& = &\left|\psi\right\rangle\left\langle \psi\right| \\ \nonumber
\hat{B}_{2}& = &\left|\psi_{\bot}\right\rangle\left\langle \psi_{\bot}\right|,
\label{projectors}
\end{eqnarray}

\noindent where

\begin{eqnarray}
\left|\psi\right\rangle& = &\rm{cos}\theta\left|0\right\rangle+e^{i\phi}\rm{sin}\theta\left|1\right\rangle \\ \nonumber
\left|\psi_{\bot}\right\rangle& = &-\rm{sin}\theta\left|0\right\rangle+e^{i\phi}\rm{cos}\theta\left|1\right\rangle.
\label{psis}
\end{eqnarray}

\noindent The resultant of the density operator when such measurements are performed on subsytem B is:

\begin{equation}
\hat{\rho}_{k}=\frac{1}{p_{k}}\left(\hat{I}\otimes \hat{B}_{k}\right)\hat{\rho}\left(\hat{I}\otimes \hat{B}_{k}\right),
\label{rhok}
\end{equation}

\noindent To obtain the final form of Eq.(\ref{Crhok}), we numerically search the $\theta$ and $\phi$ space for the set of values that maximizes Eq.(\ref{Irhok}). For a given density matrix $\hat{\rho}$, the EoF is given in terms of concurrence $\mathcal{C}\left(\hat{\rho}\right)$ by the formula \cite{Wootters}:

\begin{equation}
E\left(\hat{\rho}\right)=h\left(\frac{1+\sqrt{1-\mathcal{C}^{2}\left(\hat{\rho}\right)}}{2}\right),
\label{CEOF}
\end{equation}

\noindent where $\mathcal{C}=\rm{max}=\left\{\sqrt{\lambda_{1}}-\sqrt{\lambda_{2}}-\sqrt{\lambda_{3}}-\sqrt{\lambda_{4}}\right\}$, where ${\lambda_{i}}$ are the eigenvalues, in decreasing order, of the matrix $\hat{R}=\hat{\rho}\left(\hat{\sigma}_{y}\otimes\hat{\sigma}_{y}\right)\hat{\rho}^{T}\left(\sigma_{y}\otimes\sigma_{y}\right)$. 

For a comparison of discord with entropy, we calculate the linear entropy, defined as follows:

\begin{equation}
S_{L}=\frac{4}{3}\left(1-\rm{Tr}\left(\hat{\rho}^{2}\right)\right),
\label{S}
\end{equation}

\begin{figure}[!b]
\vspace{-3mm}
\includegraphics[angle=270,width=0.9\columnwidth]{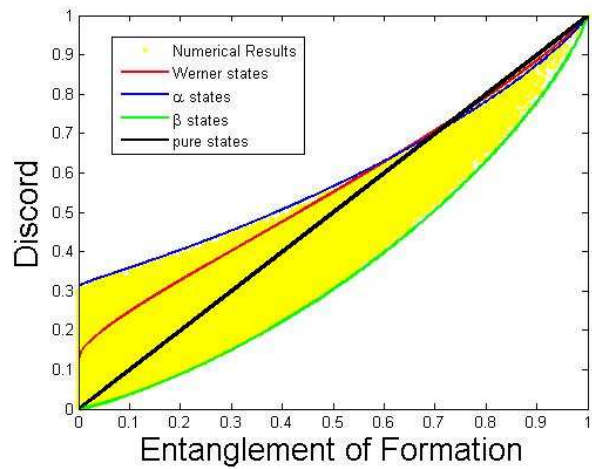}
\vspace{-2mm}
\caption{\textbf{ (Color Online) The Discord-Entanglement Horn.} 
Discord increases as entanglement increases. In the case of pure states, the two quantities are identical. While in the mixed state case the relationship broadens. However, notice that this relationship narrows in the high quantum correlated regime and is the broadest in the low correlation regime. This gives the plot its `horn' shape. The upper bound of this relationship is given by the $\alpha$-states eq.(\ref{rhoalpha}) (for $0 \leq EoF \leq 0.620$, and $0 \leq Q \leq 0.644$), the Werner states \cite{Werner} (for $0.620 \leq EoF \leq 0.746$, and $0.644 \leq Q \leq 0.746$), and the pure states (for $0.740 \leq EoF, Q \leq 1$). The lower bound is given by the $\beta$-states eq.(\ref{rhobeta}).}
\label{fig:figure1}
\end{figure}

\begin{figure}[!b]
\vspace{-3mm}
\includegraphics[angle=270,width=0.9\columnwidth]{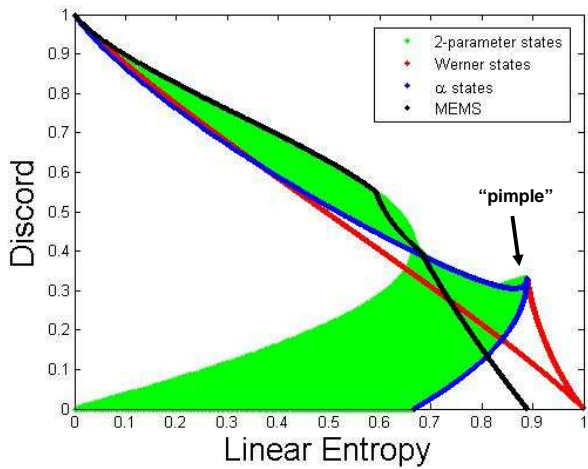}
\vspace{-2mm}
\caption{\textbf{(Color Online) The Boundaries on the Relationship Between Discord and Linear Entropy.} To illustrate easily the boundaries, this plot only includes the states that are involved in defining them. The two-parameter states, eq.(\ref{rhoab}), bound the curve from above upto $Q=1/3$ and $S_{L}=8/9$, after which the Werner states take over. Discord and Linear Entropy, as expected, display an inverse relationship: more randomeness implies less quantum correlations. One of the interesting phenomena occurs at the `pimple', where an increase in entropy results in an increase in discord. This is also the point that defines the value of linear entropy after which no entanglement can exist in the system (See \cite{Munro}). Unlike entanglement, states exist that are very close to the maximally mixed states, but still have non-zero disord. In fact the only value for entropy such that discord cannot be finite is for it being equal to 1, in the case when the system is maximally mixed.}
\label{fig:figure2}
\end{figure}

\noindent See Fig.\ref{fig:figure1} and \ref{fig:figure2} for the plot of the results. As noted above, in the case of pure states, discord and entanglement are \emph{identical}. For mixed states, the two quantities are loosly related; generally speaking the higher the entanglement, the higher the discord. The region with high quantum correlations has a narrower relationship than the one in a lower correlation regime. This results in a plot that resembles a horn (See Fig.\ref{fig:figure1}). The difference between discord and entanglement that arises in the mixed state case is due to the optimization that was done to extend the pure state case to the mixed state case. The minimization that is done over the pure state decomposition in defining entanglement gives a more pessimistic measure for quantum correlations than the maximization over all possible projectors on subsystem B that is done in defining discord for mixed states. Discord and entanglement, therefore, are \emph{not} different quantum correlations. They are two different ways to quantify these correlations; the way discord is defined seems to be capturing more of the correlations than the way entanglement is defined.

The upper bound for the discord-entanglement plot is given for the most part by two classes of MMMS: the $\alpha$-states eq.(\ref{rhoalpha}) and the Werner states eq.(\ref{Werner}). In the highly correlated regime it is bound by the pure states. The lower bound is given by another class of MMMS: the $\beta$-states eq.(\ref{rhobeta}). The Werner states, the $\alpha$-states, and the $\beta$-states are given, respectively, as follows:

\begin{equation}
\hat{\rho}(\xi)=(1-\xi)\frac{I}{4}+\xi\left|\psi^{-}\right\rangle\left\langle \psi^{-}\right|,
\label{Werner}
\end{equation}

\noindent where $-\frac{1}{3}\leq\xi\leq1$ and $\left|\psi^{-}\right\rangle=\frac{1}{\sqrt{2}}\left(\left|01\right\rangle-\left|10\right\rangle\right)$,

\begin{equation}
\hat{\rho}(\alpha)=
\left(
\begin{array}{cccc} \frac{\alpha}{2} & 0 & 0 & \frac{\alpha}{2}
\\             	   	0 & \frac{\left(1-\alpha\right)}{2} & 0 & 0  
\\									0 & 0 & \frac{\left(1-\alpha\right)}{2} & 0  
\\									\frac{\alpha}{2} & 0 & 0 & \frac{\alpha}{2} 
\end{array}
\right ),
\label{rhoalpha}
\end{equation}

\noindent where $0\leq\alpha\leq1$, and

\begin{equation}
\hat{\rho}(\beta)=
\left(
\begin{array}{cccc} \frac{\beta}{2} & 0 & 0 & \frac{\beta}{2}
\\             	   	0 & \frac{\left(1-\beta\right)}{2} & \frac{\left(1-\beta\right)}{2} & 0  
\\									0 & \frac{\left(1-\beta\right)}{2} & \frac{\left(1-\beta\right)}{2} & 0  
\\									\frac{\beta}{2} & 0 & 0 & \frac{\beta}{2} 
\end{array}
\right ),
\label{rhobeta}
\end{equation}

\noindent where $0\leq\beta\leq1$. The analytic result for quantum discord in these cases can easily be obtained from the general expression for the MMMS in \cite{Luo}. For the $\alpha$- and $\beta$-states, it is given, respectively, as follows:

\begin{eqnarray}
Q(\alpha,\zeta)&&=\left(1-\alpha\right)\rm{log}_{2}\left(1-\alpha\right)+\alpha \rm{log}_{2}\left(\alpha\right)+\left(1+\alpha\right) \nonumber \\ 
&&-(1-\zeta) (\rm{log}_{2}\left(1-\zeta\right))/2 \nonumber \\
&& -(1+\zeta) (\rm{log}_{2}\left(1+\zeta\right))/2, 
\label{Qalpha} 
\end{eqnarray} 
 
\noindent where $\zeta=\rm{max}\left\{|\alpha|, |2 \alpha-1|\right\}$, and
 
\begin{equation} 
Q(\beta)=\beta \rm{log}_{2}\left(\beta\right)+\left(1-\beta\right)\rm{log}_{2}\left(1-\beta\right)+1.
\label{Qbeta} 
\end{equation} 
 
\noindent Since these states also fall under the class of X-states, expressions for their concurrence, from which the EoF is calculated, can be found in \cite{AsmaDaniel}, for example. The concurrence of the $\alpha$- and $\beta$-states, is given, respectively, by $C(\alpha)=\rm{max}\left\{0,2\alpha-1\right\}$ and $C(\beta)=\left|2\beta-1\right|$. To prove that theses are indeed the boundaries, we performed two numeric calculations. First, we generated $10^{6}$ random density matrices to find that the relationship between their discord and entropy all fall within these bounds. The algorithm involved creating a complex and random matrix $\hat{T}$, and from it obtaining a well-behaved density matrix $\hat{\rho}$ by the relation $\hat{\rho}=\hat{T}\hat{T}^{\dagger}/\rm{Tr}\{\hat{\emph{T}}\hat{\emph{T}}^{\dagger}\}$. We also generated $10^{5}$ points very close to the vicinity of each of the boundaries, with the result that none of the points fell outside the bounds imposed by them.

In Fig.\ref{fig:figure2}, where the plot shows how much discord can be present in the system for a given amount of mixture, the boundaries are given by a different set of states. Beyond linear entropy being $8/9$, it is bound from above by the Werner states. The rest of the plot is bound from above by a class of two-parameter density matrices described as follows:

\begin{equation}
\hat{\rho}(a, b)=\frac{1}{2}
\left(
\begin{array}{cccc} a & 0 & 0 & a
\\             	   	0 & 1-a-b & 0 & 0  
\\									0 & 0 & 1-a+b & 0  
\\									a & 0 & 0 & a
\end{array}
\right ),
\label{rhoab}
\end{equation}

\noindent where $0 \leq a \leq 1$, and $a-1 \leq b \leq 1-a$. We find the analytic result for discord in this case to be:

\begin{equation}
Q(a,b)=\rm{min}\left\{a, q \right\},
\label{disAB}
\end{equation}
where

\begin{eqnarray}
q &&= -\frac{b}{2}\rm{log}_{2}\left[\frac{(1+\emph{b})(1-\emph{a}-\emph{b})}{(1-\emph{b})(1-\emph{a}+\emph{b})}\right]\nonumber \\
&& + \frac{\emph{a}}{2}\rm{log}_{2}\left[\frac{4\emph{a}^{2}}{(1-\emph{a})^{2}-\emph{b}^{2}}\right] \nonumber \\ 
&&-\frac{\sqrt{\emph{a}^{2}+\emph{b}^{2}}}{2}\rm{log}_{2}\left[\frac{1+\sqrt{\emph{a}^{2}+\emph{b}^{2}}}{1-\sqrt{\emph{a}^{2}+\emph{b}^{2}}}\right] \nonumber \\
&& + \frac{1}{2}\rm{log}_{2}\left[\frac{4((1-\emph{a})^{2}-\emph{b}^{2})}{(1-\emph{b}^{2})(1-\emph{a}^{2}-\emph{b}^{2})}\right], \nonumber \\
\label{Q2}
\end{eqnarray}

\noindent and the expression for cuncurrence in this case is $\mathcal{C}(a,b)=\rm{max}\{0,|\emph{a}|-\sqrt{(1-\emph{a})^{2}-\emph{b}^{2}}\}$.

As can be seen in Fig.\ref{fig:figure2}, the Maximally Entangled Mixed States (MEMS) \cite{Munro}, and the $\alpha$-states, both of which fall under this class of states, bound the plot at different regions. There is no lower bound to the relationship between discord and entropy, as the area below the two-parameter states as well as the Werner states gets filled up when the whole range of density matrices is included. Also note that the case when $a=q$ is what gives the bounding line that slopes down from the `pimple'. Again, as is the case with discord and EoF, to verify these boundaries, points representing $10^{6}$ random density matrices were generated, as well as $10^{5}$ points in the near vicinity of these bounds. Other points to note about the figure is, first, that at the \emph{pimple}, after which no entanglement can exist in the system \cite{Munro}, a rise in linear entropy results in a rise in discord. The states in this region are interesting to investigate, since experimentally speaking, more noise in the system at that stage enhances the quantum correlations. Also, unlike the case with entanglement, even for cases where linear entropy is arbitrary close to the maximally mixed states at unit entropy, discord can still be finite. It is, therefore, not surprising that for states in which ESD occurs, similar behaviour is not observed for discord \cite{SDQ,Ferraro}.

In conclusion, this work describes the relationship between discord and entanglement for the general two-level bipartite system. We have shown that in the general case of mixed states the two quantum correlations vary, with the relationship broadening in the low quantum regime. We conclude that they describe the same set of quantum correlations, as can be seen in the pure state case, and although they vary in the case of mixed states, this is due to different methods of optimization used to extend the correlations from the pure state case to the mixed state case. The way discord is defined happens to capture more of the correlations than the way entanglement, in the form of concurrence, is defined. Some questions that arises are: What is the optimal method to quantify quantum correlations? What is the meaning of the values for discord and EoF that are between the extreme cases of $0$ and $1$? We also reveal the general relationship between discord and linear entropy highlighting interesting differences with a similar analysis done for entanglement \cite{Munro}: at the point where entanglement disappears from the system, discord increases in value, and discord can be nonzero unless linear entropy is identically equal to one.
\ 

\ 
 
\noindent This work was supported by NSERC.

\end{document}